\documentstyle[12pt,epsfig]{article}

\newlength{\dinwidth}
\newlength{\dinmargin}
\begin{document}
\def\bold#1{\setbox0=\hbox{$#1$}%
     \kern-.025em\copy0\kern-\wd0
     \kern.05em\%\baselineskip=18ptemptcopy0\kern-\wd0
     \kern-.025em\raise.0433em\box0 }
\def\slash#1{\setbox0=\hbox{$#1$}#1\hskip-\wd0\dimen0=5pt\advance
         to\wd0{\hss\sl/\/\hss}}
\newcommand{\be}{\begin{equation}}
\newcommand{\ee}{\end{equation}}
\newcommand{\bea}{\begin{eqnarray}}
\newcommand{\eea}{\end{eqnarray}}
\newcommand{\nn}{\nonumber}
\newcommand{\dd}{\displaystyle}
\newcommand{\bra}[1]{\left\langle #1 \right|}
\newcommand{\ket}[1]{\left| #1 \right\rangle}
\newcommand{\spur}[1]{\not\! #1 \,}
\thispagestyle{empty} \vspace*{1cm}
\rightline{BARI-TH/526-05}\vspace*{2cm}
\begin{center}
  \begin{LARGE}
  \begin{bf}
Bounding effective parameters   \\ \vspace*{0.4cm}
 in the chiral Lagrangian  for  excited \\ \vspace*{0.4cm}
  heavy mesons \\ \vspace*{0.4cm}
  \end{bf}
  \end{LARGE}
\end{center}
\vspace*{8mm}
\begin{center}
\begin{large}
P.  Colangelo$^a$, F. De Fazio$^a$ and R. Ferrandes$^{a,b}$
\end{large}
\end{center}
\begin{center}
\begin{it}
$^a$Istituto Nazionale di Fisica Nucleare, Sezione di Bari,
Italy\\ $^b$ Universit\'a degli Studi di Bari, Italy
\end{it}
\end{center}
\begin{quotation}
\vspace*{1.5cm}
\begin{center}
  \begin{bf}
  Abstract\\\end{bf}
\end{center}
\vspace*{0.5cm} \noindent We use recent experimental data on
charmed mesons to constrain  three  coupling constants in the
effective  lagrangian  describing the interactions of excited
heavy-light mesons with light pseudoscalar mesons  at order
$m_Q^{-1}$. Predictions in the beauty sector are also derived.

\end{quotation}
\newpage
\baselineskip=18pt \vspace{2cm} \noindent

The coupling constants and  the mass parameters in  effective Lagrangians which reproduce  QCD
in specific limits represent important  input parameters for the
description of  the hadron processes. Therefore  their determination, either by theoretical
approaches or  by  phenomenological analyses,  is  relevant for the use of the related effective theory.
The case of the effective
Lagrangian describing the strong interactions of heavy-light hadrons with the octet of
pseudo Goldstone bosons is not an exception, and it is noticeable that data recently collected
at the $B$ factories and at the Fermilab Tevatron can  constrain a few of such parameters, thus
allowing to  exploit this theoretical framework to make, for example,  predictions that
can be tested at the new experiments. This note is devoted to such a discussion.

\vspace*{1cm}
The  heavy quark chiral effective theory
is constructed  starting from the spin-flavour symmetry occurring in QCD
for hadrons comprising a single heavy quark, in the infinite heavy quark mass limit,
and from the chiral symmetry valid in the massless limit for the light quarks
\cite{hqet_chir}.
The heavy quark spin-flavour symmetry allows to classify heavy $Q \bar q$ mesons into
doublets labeled by the value of the
 angular momentum $s_\ell$ of the light degrees of freedom: $s_\ell=s_{\bar q}+ \ell$, $s_{\bar q}$
being the light antiquark spin and $\ell$ the orbital angular
momentum of the light degrees of freedom relative to the heavy
quark \cite{positivep}.
The lowest lying $Q \bar q$ mesons correspond to $\ell=0$, then
$ s_\ell^P={1 \over 2}^-$;
 this  doublet comprises  two states with spin-parity
$J^P=(0^-,1^-)$:
$P=D_{(s)} (B_{(s)})$  and  $P^{*}=D^*_{(s)} (B^*_{(s)}) $
mesons in case of charm (beauty) heavy quark, respectively.
 For $\ell=1$ it could be either $ s_\ell^P={1 \over 2}^+$ or $
s_\ell^P={3 \over 2}^+$.  The  two corresponding doublets have
$J^P=(0^+,1^+)$ and $J^P=(1^+,2^+)$.  We denote the members of the
$J^P_{s_\ell}=(0^+,1^+)_{{1/2}}$ doublet as $(P^*_{0},P_{1}^\prime)$
 and those of the
$J^P_{s_\ell}=(1^+,2^+)_{3/2}$  doublet as $(P_{1},P^*_{2})$, with
$P=D,D_s,B,B_s$.
The negative and positive  parity doublets can be respectively described by the  fields
$H_a$, $S_a$ and  $T_a^{\mu}$, $a=u,d,s$ being a light flavour index:
\bea
H_a & =& \frac{1+{\rlap{v}/}}{2}[P_{a\mu}^*\gamma^\mu-P_a\gamma_5]  \label{neg} \\
S_a &=& \frac{1+{\rlap{v}/}}{2} \left[P_1^{\prime \mu}\gamma_\mu\gamma_5-P_0^*\right]  \label{pos1}\\
T_a^\mu &=&\frac{1+{\rlap{v}/}}{2} \left\{ P^{*\mu\nu}_{2a} \gamma_\nu -P_{1a\nu} \sqrt{3 \over 2} \gamma_5 \left[
g^{\mu \nu}-{1 \over 3} \gamma^\nu (\gamma^\mu-v^\mu) \right]
\right\} \;, \label{pos2}
\eea
with the various operators annihilating mesons of four-velocity $v$
which is conserved in  strong interaction processes. The heavy field operators  contain a
factor $\sqrt{m_P}$ and have dimension $3/2$.

The octet of light pseudoscalar mesons can be introduced using the
 representation $\displaystyle \xi=e^{i {\cal M} \over
f_\pi}$ and $\Sigma=\xi^2$; the matrix ${\cal M}$ contains
$\pi, K$ and $\eta$ fields:
\begin{equation}
{\cal M}= \left(\begin{array}{ccc}
\sqrt{\frac{1}{2}}\pi^0+\sqrt{\frac{1}{6}}\eta & \pi^+ & K^+\nonumber\\
\pi^- & -\sqrt{\frac{1}{2}}\pi^0+\sqrt{\frac{1}{6}}\eta & K^0\\
K^- & {\bar K}^0 &-\sqrt{\frac{2}{3}}\eta
\end{array}\right)
\end{equation}
with $f_{\pi}=132 \; $ MeV.

The effective QCD Lagrangian is constructed imposing
invariance under heavy quark spin-flavour transformations and chiral transformations.
The kinetic term
 \begin{eqnarray} {\cal L} &=& i\; Tr\{ {\bar H}_b v^\mu
D_{\mu ba}  H_a \}  + \frac{f_\pi^2}{8}
Tr\{\partial^\mu\Sigma\partial_\mu \Sigma^\dagger \} \nn \\ &+&
Tr\{ {\bar S}_b \;( i \; v^\mu D_{\mu ba} \; - \; \delta_{ba} \;
\Delta_S)
 S_a \}
+   Tr\{ {\bar T}_b^\mu \;( i \; v^\mu D_{\mu ba} \; - \;
\delta_{ba} \; \Delta_T)  T_{a \mu} \}   \label{L}
\end{eqnarray}
involves the operators $D$ and $\cal A$:
\begin{eqnarray}
D_{\mu ba}&=&-\delta_{ba}\partial_\mu+{\cal V}_{\mu ba}
=-\delta_{ba}\partial_\mu+\frac{1}{2}\left(\xi^\dagger\partial_\mu
\xi
+\xi\partial_\mu \xi^\dagger\right)_{ba}\\
{\cal A}_{\mu ba}&=&\frac{i}{2}\left(\xi^\dagger\partial_\mu
\xi-\xi
\partial_\mu \xi^\dagger\right)_{ba} \; ,
\end{eqnarray}
and the mass parameters $\Delta_S$ and $\Delta_T$ which
represent the mass splittings between   positive and
negative parity doublets. They can be expressed
 in terms of the spin-averaged masses:
 $\Delta_S= \overline M_S - \overline M_H$ and  $\Delta_T= \overline M_T - \overline M_H$
with
\bea {\overline M}_H& =& {3 M_{P^*}+M_P  \over 4} \nn \\
{\overline M}_S &=& {3 M_{P^\prime_1}+M_{P_0^*} \over 4}  \\
{\overline M}_T &=& {5 M_{P^*_2}+3M_{P_1} \over 8} \,\, .\nn  \label{lambdapar}\eea

At the leading order in the heavy quark expansion
the decays  $H \to H^{\prime} M$, $S\to H^{\prime} M$ and
$T\to H^{\prime} M$  ($M$ a light pseudoscalar meson) are described by
the lagrangian terms:
\bea
{\cal L}_H &=& \,  g \, Tr [{\bar H}_a H_b \gamma_\mu \gamma_5 {\cal
A}_{ba}^\mu ] \nn \\
{\cal L}_S &=& \,  h \, Tr [{\bar H}_a S_b \gamma_\mu \gamma_5 {\cal
A}_{ba}^\mu ]\, + \, h.c. \nn \\
{\cal L}_T &=&  {h^\prime \over \Lambda_\chi}Tr[{\bar H}_a T^\mu_b
(i D_\mu {\spur {\cal A}}+i{\spur D} { \cal A}_\mu)_{ba} \gamma_5
] + h.c. \label{lag-hprimo}\eea
where $\Lambda_\chi$ is  a chiral symmetry-breaking scale; we  use
$\Lambda_\chi = 1 \, $ GeV.
${\cal L}_S$ and ${\cal L}_T$ describe transitions of positive parity heavy mesons with
the emission of light pseudoscalars in $S$ and $D$ wave, respectively.
The coupling constants  $h$ and $h^\prime$  weight the interactions of
$S$ and $T$  heavy-light mesons with the light pseudoscalar mesons.

Corrections to the heavy quark limit  induce symmetry
breaking terms  suppressed by increasing powers of
$m_Q^{-1}$ \cite{Falk:1995th}. Mass degeneracy between the members
of the  meson doublets is broken by the  terms:
\be
{\cal L}_{1/m_{Q}}={1 \over 2 m_{Q}} \left\{ \lambda_H Tr [{\bar H}_{a}
\sigma^{\mu \nu} H_{a} \sigma_{\mu \nu}]-\lambda_S Tr [{\bar S}_{a}
\sigma^{\mu \nu} S_{a} \sigma_{\mu \nu}]+\lambda_T Tr [{\bar T}^\alpha_{a}
\sigma^{\mu \nu} T^\alpha_{a} \sigma_{\mu \nu}] \right\} \,\,\,\,
\label{mass-viol} \ee
where the constants $\lambda_H$, $\lambda_S$ and $\lambda_T$ are related  to
the hyperfine mass splittings:
\bea \lambda_H &=& {1 \over 8} \left( M_{P^*}^2-M_P^2
\right) \nn \\\lambda_S &=& {1 \over 8} \left(
M_{P^\prime_1}^2-M_{P_0^*}^2 \right)  \\\lambda_T &=& {3 \over
8} \left( M_{P^*_2}^2-M_{P_1}^2 \right) \,\,\, . \nn \label{lambdas}\eea

Other two  effects stemming from spin
symmetry-breaking terms concern the possibility that  the members of the
$s_\ell={3\over 2}^+$ doublet can also decay in S wave into the lowest lying heavy
mesons and pseudoscalars, and that a
mixing may be induced between the two $1^+$ states belonging to
the two positive parity  doublets with different $s_\ell$. The
corresponding terms in the effective Lagrangian are:
\bea
{\cal L}_{D_1}&=&{f \over 2m_Q \Lambda_\chi}Tr [{\bar H}_{a} \sigma^{\mu \nu}
T^\alpha_{b} \sigma_{\mu \nu} \gamma^\theta \gamma_5 (i D_\alpha {\cal
A}_\theta+ iD_\theta {\cal A}_\alpha)_{ba}] + h.c.  \, \label{ld1} \\
{\cal L}_{mix}&=&{g_1 \over 2m_Q} Tr[{\bar S}_{a} \sigma^{\mu
\nu}T_{\mu a} \sigma_{ \nu \alpha}v^\alpha] \, + h.c. \label{lmix}\eea
 Notice that ${\cal L}_{D_1}$ describes both $S$ and $D$ wave decays.
The mixing angle between the two $1^+$ states:
\bea \ket{P_1^{phys}}&=&\cos \theta
\ket{P_1}+ \sin{\theta} \ket{P_1^\prime} \label{d1phys} \\
\ket{P_1^{\prime phys}}&=&-\sin \theta
\ket{P_1}+ \cos{\theta} \ket{P_1^\prime} \label{d1primephys}
\eea
 can be related to the coupling constant $g_1$ and to the mass splitting:
\be
\tan \theta={\sqrt{\delta^2+\delta_g^2}-\delta \over \delta_g}
\label{mix-angle} \ee
 where
$\delta=\displaystyle{\Delta_T -\Delta_S \over 2}$ and
$\delta_g=-\displaystyle{\sqrt{2 \over 3}{g_1 \over m_Q}}$.

\vspace*{1cm}
The parameters  in the
various terms of the effective Lagrangian are
universal and their determination  is important in the definition of the effective theory and
 in the applications to the hadron phenomenology. Data recently collected on
 charmed and charmed-strange mesons, together with information on previously known
 positive parity charmed states, allow us to determine some of them
 by an analysis   previously  impossible  due to the lack of enough experimental input.

A new result is the observation of  charmed mesons which can be accomodated
in the    $s_\ell^{P}={1\over 2}^{+}$ doublet.
Two broad states  which could be identified as the
$D^*_0$ and  $D^\prime_1$ mesons have been  observed by Belle
\cite{Abe:2003zm},  FOCUS \cite{Link:2003bd}  and CLEO \cite{Anderson:1999wn} Collaborations.
The masses and widths measured by  Belle  (the only experiment
which separately observes the two states) are:
 $M_{D^{*0}_0}=2308 \pm 17\pm 15\pm28$
MeV, $\Gamma(D^{*0}_0)=276 \pm 21\pm 18\pm 60$ MeV and
$M_{D^{\prime 0}_1}=2427 \pm 26 \pm 10\pm15$
MeV, $\Gamma(D^{\prime 0}_1)=384^{+107}_{-75} \pm 24 \pm 70$ MeV, while
the average  values from the various experiments are: $M_{D^*_0}=2351 \pm 27$
MeV, $\Gamma(D^*_0)=262 \pm 51$ MeV (from Belle and FOCUS),
and  $M_{D^\prime_1}=2438 \pm 30$
MeV, $\Gamma(D^\prime_1)=329 \pm 84$ MeV (from Belle and CLEO).
In the  charm-strange sector  the two mesons $D^{*}_{sJ}(2317)$
 and $D_{sJ}(2460)$ \cite{Aubert:2003fg}
 naturally fit in the doublet ($D^*_{s0}$, $D^\prime_{s1}$).  Being below the  $DK$ and $D^*K$ decay thresholds, respectively,
 they are narrow  \cite{Colangelo:2004vu}.

The two sets of measurements  allow  to determine a few  parameters  in
eqs.(\ref{L}), (\ref{mass-viol}), (\ref{ld1}) and (\ref{lmix}).
In Table \ref{lambdaHST} we collect the values of
$\lambda_H$, $\lambda_S$ and $\lambda_T$ obtained
 using the masses of the charmed and beauty states
reported by PDG \cite{PDG}  with two exceptions.
The first one  is $\lambda_S$ in case of non-strange charmed mesons,  which we derive from the
Belle measurement.  Had we used mass values averaged over Belle, CLEO
and Focus measurements we would have obtained a smaller value for  $\lambda_{S}$,
 compatible  within the uncertainties with the value in Table  \ref{lambdaHST}.
 The second exception concerns $B^*_s$, reported by PDG in the list of particles needing confirmation with
  $m_{B^*_s}=5416 \pm 3.5 \ {\rm MeV}$; we use the mass recently measured by the CLEO Collaboration:
 $m_{B^*_s}=5414 \pm 1 \pm 3  \ {\rm MeV}$  \cite{Bonvicini:2005ci}.
 In Table  \ref{lambdaHST} we also report  the  spin averaged masses and  the mass splitting
 between positive and negative doublets.
\begin{table}[h]
    \caption{ $\lambda_{i}$ parameters    obtained using data in PDG \cite{PDG}.
  For the determination of $\lambda_S$   in case of $c \bar q$ see  text. The  spin-averaged masses
  for the various doublets and  the mass splittings $\Delta_{S}$ 
and $\Delta_{T}$ are also reported.}
    \label{lambdaHST}
    \begin{center}
    \begin{tabular}{|lcccc|}
      \hline \hline
  & $c{\bar q}$  & $c{\bar s}$ &$b{\bar q}$ &$b{\bar s}$\\ \hline $\lambda_H$ & $ (261.1 \pm 0.7\,\, {\rm MeV})^2$ &
 $ (270.8 \pm 0.8 \,\, {\rm MeV})^2$
 & $ (247 \pm 2 \,\, {\rm MeV})^2$ & $ (252 \pm 10 \,\, {\rm MeV})^2$ \\
$\lambda_S$ & $ (265 \pm 57 \,\, {\rm MeV})^2$ &$(291 \pm 2 \,\,
{\rm MeV})^2$
 &  & \\
$\lambda_T$ & $(259\pm 10 \,\, {\rm MeV})^2$ &$ (266 \pm 6\,\,
{\rm MeV})^2$ & &\\  \hline
$\overline M_{H}$ & $1974.8\pm 0.4 \, {\rm MeV}$ & $2076.1\pm 0.5 \, {\rm MeV}$ & $5313.5\pm 0.5 \, {\rm MeV}$ & $5404\pm 3 \, {\rm MeV}$ \\
$\overline M_{S}$ & $2397\pm 28 \, {\rm MeV}$ & $2424\pm 1 \, {\rm MeV}$ & & \\
$\overline M_{T}$ & $2445.1\pm 1.4 \, {\rm MeV}$ & $2558 \pm 1 \, {\rm MeV}$ & & \\ \hline
$ \Delta_{S}$ & $422\pm 28 \, {\rm MeV}$ & $348\pm 1 \, {\rm MeV}$ & & \\
$ \Delta_{T}$ & $470.3\pm 1.5 \, {\rm MeV}$ & $482\pm 1 \, {\rm MeV}$ & & \\
 \hline \hline
    \end{tabular}
  \end{center}
\end{table}

A few considerations are in order. First,  observable SU(3) effects appear in  charm determinations of  $\lambda_{H}$,
while analogous effects  are not evident in  $\lambda_{{S,T}}$ due to the larger  experimental uncertainties.
Second, a sizeable heavy quark mass effect remains in  $\lambda_{H}$ when it is determined from charm and beauty data,
meaning that further terms in the  heavy quark mass expansion should be considered for describing data at the present level of accuracy.
Finally,
  $\lambda_H$ and $\lambda_S$ are not equal, at odds with the relation
 $\lambda_H \simeq \lambda_S$  suggested by a description of negative and positive $s_{\ell}= {1 \over 2}$ states
as chiral partners   \cite{Bardeen:2003kt}
\footnote{The idea of chiral doubling as a consequence of the restoration of the
chiral symmetry has been recently challenged  in \cite{Jaffe:2005sq},
where it is argued that when the symmetry is broken in the Nambu-Goldstone mode
through the appearance of pions, it cannot be manifested somewhere in the spectrum.}.

Since the  parameters $\lambda_H$, $\lambda_S$, $\lambda_T$ are independent of the heavy quark mass,   they
can be used to constrain the masses of excited beauty mesons.  We impose that the
splittings  $\Delta_S$ and $\Delta_T$ are the same for charm and beauty: this is true in the rigorous heavy quark limit,
while at $O(1/m_{Q})$ such an assertion corresponds to assuming that the matrix element of the kinetic energy operator
is the same for the three doublets.  Furthermore,
 SU(3)$_F$ effects are included in the determinations based on the charm-strange sector.  The results are reported in Table \ref{bmasses}.
  It is worth noticing that
the $B_{s0}^*$, $B_{s1}^\prime$ masses are below the $B K$ and $B^{*} K$ thresholds,
and therefore they  are expected to  be  narrow (as also argued  in
\cite{Colangelo:2003vg,Colangelo:2004vu}). The search for such narrow resonances is in the physics programmes of  collider experiments,
the Tevatron and the LHC.

\begin{table}[ht]
    \caption{Predicted masses of excited beauty mesons.}
    \label{bmasses}
    \begin{center}
    \begin{tabular}{|ccccc|}
      \hline \hline
 & $ B^*_{(s)0} \, (0^{+})$ & $ B^\prime_{(s)1} \, (1^{+})$ & $ B_{(s)1} \, (1^{+})$ & $B^*_{(s)2} \,(2^{+})$ \\ \hline
$b \bar q$& $5.70\pm0.025 \,{\rm GeV}$ &$5.75\pm 0.03 \,{\rm GeV}$ & $5.774\pm 0.002 \,{\rm GeV}$ & $5.790\pm 0.002 \,{\rm GeV}$ \\
$b \bar s$& $5.71\pm 0.03 \,{\rm GeV}$ &$5.77\pm 0.03 \,{\rm GeV}$ & $5.877\pm 0.003 \,{\rm GeV}$ & $5.893\pm 0.003 \,{\rm GeV}$ \\
\hline \hline
    \end{tabular}
  \end{center}
\end{table}

In the above determinations we have neglected the mixing  angle between the
two $1^+$ states $D_1$ and $D_1^\prime$. Considering, instead,
 the  result  $\theta_c=-0.10 \pm 0.03 \pm 0.02\pm 0.02 \,\,rad$ \cite{Abe:2003zm}
 and  using $\Delta_T$ and $\Delta_S$ in Table \ref{lambdaHST}
 together with eq.(\ref{mix-angle}) and $m_{c}=1.35 \,$ GeV,
we can compute the coupling $g_1$ in (\ref{lmix}): $g_1=0.008\pm
0.006 \,$  GeV$^2$,  therefore compatible with zero. In the beauty
system, for  $m_b=4.8 \,$  GeV,  one  obtains:  $\theta_b \simeq
-0.028 \pm 0.012\,\,\, rad$.

\vspace*{1cm}
To determine the couplings $h^\prime$  and
$f$ in eqs.(\ref{lag-hprimo}-\ref{ld1})   we consider
 the widths of the two members of the     $s_\ell^P={3 \over 2}^+$
doublet,  $D_1$ and $D_2^*$, identifying
 the observed meson $D_1(2420)$  with the physical state  (\ref{d1phys}):
\bea \Gamma(D_2^{*0} \to D^+ \pi^-)&=& {4 \over 15 \pi }{m_D \over
m_{D^*_2}}{|{\vec p}_\pi|^5 \over f_\pi^2} {1 \over
\Lambda_\chi^2}
\left(h^\prime -{f \over m_c} \right)^2  \label{widthsd2a} \\
\Gamma(D_2^{*0} \to D^{*+} \pi^-)&=&{2 \over 5 \pi }{m_{D^*} \over
m_{D^*_2}}{|{\vec p}_\pi|^5 \over f_\pi^2} {1 \over
\Lambda_\chi^2} \left(h^\prime -{f \over m_c} \right)^2
\label{widthsd2} \\
 \Gamma(D_1^{0} \to D^{*+} \pi^-) &=& {2
\over 3 \pi} {m_{D^*} \over m_{D_1}}    {|{\vec p}_\pi|^5 \over
f_\pi^2} {1 \over \Lambda_\chi^2} \left[\left(h^\prime +{5 \over
3} {f \over m_c} \right)^2+{32 \over 9} {f^2 \over m_c^2} \right]
\label{widthD1} \eea
and
\be
 \Gamma(D_{1}^{\prime0} \to D^{*+} \pi^-) = {1
\over 2 \pi} {m_{D^*} \over m_{D^{\prime}_1}}    {|{\vec p}_\pi|  h^{2}\over
f_\pi^2} (m_{\pi}^2+ |{\vec p}_\pi|^{2}) \,\,\, .
\ee
Analogous equations hold for charmed-strange meson transitions:
$D_{s2}^{*+} \to D^{(*)+}K^0$, $D^{(*)0} K^+ $ and $D_{s1}^+ \to D^{*+} K^0, \,D^{*0} K^+$.
Assuming that the full widths are saturated by two-body decay modes with
single pion (kaon) emission,  we can determine $h^\prime$ and $f$ using recent
 results  from Belle Collaboration \cite{Abe:2003zm}:
\bea
\Gamma(D_2^{*0})&=& 45.6 \pm 4.4 \pm 6.5 \pm 1.6
\,\,\, {\rm MeV} \nn \\
\Gamma(D_1^{0}) &=& 23.7 \pm 2.7 \pm 0.2 \pm 4.0 \,\,\, {\rm MeV}
\label{exp-gamma} \,\,. \eea
Notice  that, while  $\Gamma(D_1^{0})$ is compatible with the  value reported
by PDG ($\Gamma(D_1^{0}) = 18.9^{+4.6}_{-3.5}$ MeV) and with a recent measurement
by CDF Collaboration \cite{CDF}, the width of $D_2^{*0}$ is
larger than the PDG value ($\Gamma(D_1^{0}) = 23\pm5$ MeV), while it is  consistent with a Focus measurement:
$\Gamma(D_2^{*0})=38.7 \pm 5.3 \pm 2.9  $ MeV \cite{Link:2003bd}
and   a CDF measurement:
$\Gamma(D_2^{*0})=49.2 \pm 2.1 \pm 1.2  $ MeV \cite{CDF}.
We use  $h=-0.56$ \cite{Colangelo:1995ph}.
%
\begin{figure}[t]
 \begin{center}
  \includegraphics[width=0.45\textwidth] {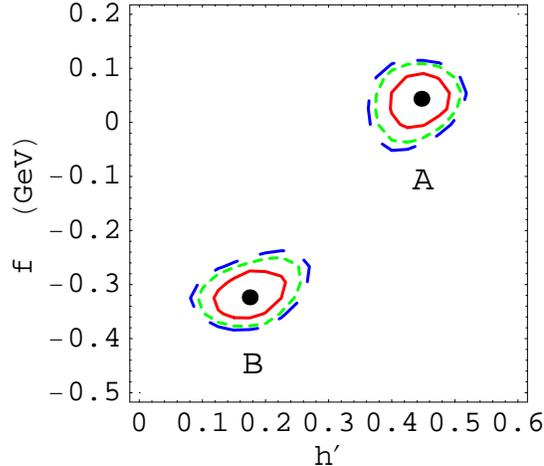}
\vspace*{0mm}
 \caption{Regions in the $(h^\prime,f)$ plane
 constrained by  the  widths of $D_2^{*0}$ and $D_1^0$. Only the region A is also compatible with
  the  constraints on the parameter $R$ in eq.(\ref{Rpar}).}
  \label{regions}
 \end{center}
\end{figure}
%
In the plane $(h^\prime,f)$ four regions are allowed by  data which, due to symmetry $(h^\prime, f) \to (-h^\prime, -f)$,
 reduce to the two inequivalent regions depicted in fig.\ref{regions}.

Notice that the two terms in square brakets in eq.(\ref{widthD1})
correspond to D-wave  and S-wave pion emission,
respectively.  A further constraint is the  Belle
 measurement of the helicity angle distribution in the
decay $D_{s1}(2536) \to D^{*+} K_S^0$,  with the determination of the ratio
 \be R= \displaystyle{\Gamma_S \over
\Gamma_S+\Gamma_D } \label{Rpar}\,\,\, ,\ee
$\Gamma_{S,D}$  being the S and D wave
partial widths, respectively \cite{Abe:2005xj}:
 $0.277 \le R \le 0.955$
(a measurement of the ratio $R$ versus the phase difference  between $S$ and $D$  was
obtained by CLEO Collaboration for non-strange mesons  \cite{Avery:1994yc}).
Although the range of $R$ is  wide,  it allows to
exclude  the  region $B$ in  fig.\ref{regions},
 leaving only the region $A$  that can be represented as
\be
h^\prime= 0.45 \pm 0.05  \hspace*{1cm}  f=0.044 \pm 0.044  \,\,{ \rm GeV}\label{results} \,\,.\ee
The coupling constant $f$ is    compatible with zero,  indicating
that the contribution of the  lagrangian term (\ref{ld1})  is small. Since also
the coupling $g_{1}$ turns out to be small, the two $1^{+}$ states corresponding to the
$s_{\ell}^{P}={1 \over 2}^{+},{3 \over 2}^{+} $
practically coincide with the physical states. For
 the   width of $D_{s1}(2536)$ we predict
\be \Gamma(D_{s1}(2536))=2.5 \pm 1.6 \,\,\,{\rm MeV} \label{2536wid} \ee
 compatible with the present  bound:
$\Gamma(D_{s1}(2536))<2.3$ MeV \cite{PDG}.

It is possible to  predict the  widths of excited $B_{(s)}$ mesons,
the   results  are collected in  Table \ref{larghezze}.
\begin{table}[h]
    \caption{Predictions for decay widths  and branching fractions of
    $J^P_{s_\ell}=(1^+,2^+)_{3 \over 2}$
     beauty mesons. The full  widths  are obtained
     assuming  saturation of the two-body modes.}
    \label{larghezze}
    \begin{center}
    \begin{tabular}{|c c c | c c c|}
      \hline \hline
      Mode & $\Gamma$(MeV) & BR & Mode & $\Gamma$(MeV) & BR  \\\hline
$B_2^{*0} \to B^+ \pi^-$ & $20 \pm 5$ & $0.34$ & $B_{s2}^{*0} \to B^+K^-$ & $4 \pm 1$ & $0.37$ \\
$B_2^{*0} \to B^0 \pi^0$  & $10.0 \pm 2.3$ & $0.17$& $B_{s2}^{*0} \to B^0 K^0$  &  $4 \pm 1$ & $0.34$\\
$B_2^{*0} \to B^{*+} \pi^-$  & $18 \pm 4$ &$0.32$ & $B_{s2}^{*0} \to B^{*+} K^-$  & $ 1.7 \pm 0.4$ & $0.15$\\
$B_2^{*0} \to B^{*0} \pi^0$  & $ 9.3 \pm 2.2$ & $0.16$& $B_{s2}^{*0} \to B^{*0} K^0$  & $1.5 \pm 0.4$ &$0.13$\\
$B_2^{*0}$ & $57.3\pm 13.5$ & &$B_{s2}^{*0}$ & $11.3\pm2.6 $ &\\
  \hline
$B_1^{0} \to B^{*+} \pi^-$  & $28 \pm 6$ & $0.66$& $B_{s1}^{0} \to B^{*+} K^-$  & $ 1.9 \pm 0.5$ &$0.54$\\
$B_1^{0} \to B^{*0} \pi^0$  & $ 14.5 \pm 3.2$ &$0.34$ & $B_{s1}^{0} \to B^{*0} K^0$ & $1.6 \pm 0.4$ &$0.46$ \\
 $B_1^{0}$ & $43\pm10$ & & $B_{s1}^{0}$ & $3.5\pm1.0$ &\\
     \hline \hline
    \end{tabular}
  \end{center}
\end{table}
Moreover, for  $B_1$ and $B_{s1}$
 the   ratios  (\ref{Rpar}) turn  out to be compatible with zero:
$R = 0.01 \pm 0.01$ (for $B_1$) and
$R = 0.1 \pm 0.1$ (for $B_{s1}$).
A word of caveat is needed here, since these predictions are obtained only considering
the heavy quark spin-symmetry breaking terms in the effective Lagrangian; corrections due to
spin-symmetric but heavy flavour breaking terms involve additional couplings for which
no sensible phenomenological information is currently available, so that they
cannot be reliably bounded. Keeping this  in mind, we notice that the full widths
of $B_{(s)2}^{*0}$ and   $B_{(s)1}^{0}$ are determined with remarkable accuracy.
As for the present experimental information concerning these states,
PDG reports in the listing of states needing confirmation a
$B^{*}_{J}(5732)$ signal, which could be considered as stemming from several
narrow and broad resonances,
 with (average)  mass $5698\pm8 \,$ MeV and (average) width of
 $128 \pm 18 \,$ MeV
 \cite{PDG}.
 The separation of this signal in its components could be done using our predictions.
A $B^{*}_{sJ}(5850)$ signal is also reported
 with   mass $5853\pm 16 \,$ MeV and width of $47 \pm 22 \, $ MeV \cite{PDG},
within our predictions.

\vspace*{1cm} In conclusion, we have exploited recent observations
and measurements concerning excited charm mesons to determine two
coupling constants governing their strong decays to light
pseudoscalar mesons, as well as the mixing parameter between the
two $J^P=1^+$ states. Furthermore, we have estimated masses and
decay rates of corresponding beauty states, and these predictions
will be checked at the future experimental environments, such as
the LHC, where such states could be observed.


\newpage
\begin{thebibliography}{99}

\bibitem{hqet_chir}
M.B.Wise, Phys. Rev. {\bf D 45}  (1992)  R2188;
G.Burdman and J.F.Donoghue, Phys. Lett. {\bf B 280} (1992) 287;
P.Cho, Phys. Lett. {\bf B 285} (1992)  145;
H.-Y.Cheng {\it et al.,}  Phys. Rev. {\bf D 46} (1992)  1148;
R.Casalbuoni {\it et al.,} Phys. Lett. {\bf B 299} (1993) 139.

\bibitem{positivep}
N.Isgur and M.B.Wise, Phys. Rev. Lett. {\bf 66} (1991) 1130; Phys. Rev. {\bf D 43} (1991) 819;
U.Kilian, J.G.K\"orner and D.Pirjol, Phys. Lett. {\bf B 288} (1992) 360;
A.F.Falk and M.Luke, Phys. Lett. {\bf B 292} (1992) 119.


\bibitem{Falk:1995th}
  A.~F.~Falk and T.~Mehen,
  Phys.\ Rev.\ D {\bf 53} (1996) 231.

\bibitem{Abe:2003zm}
  K.~Abe {\it et al.}  [Belle Collaboration],
  Phys.\ Rev.\ D {\bf 69} (2004) 112002.

\bibitem{Link:2003bd}
J.~M.~Link {\it et al.}  [FOCUS Collaboration],
Phys. Lett. {\bf B586}  (2004) 11.

\bibitem{Anderson:1999wn}
S.~Anderson {\it et al.}  [CLEO Collaboration],
Nucl. Phys.  {\bf A663} (2000) 647.

\bibitem{Aubert:2003fg}
B.~Aubert {\it et al.}  [BABAR Collaboration],
Phys. Rev. Lett.  {\bf 90} (2003) 242001;
Y.~Mikami {\it et al.} [Belle Collaboration], Phys. Rev. Lett.
{\bf 92} (2004) 012002;
K.~Abe {\it et al.},
Phys.\ Rev.\ Lett.\  {\bf 92} (2004) 012002;
%
D.~Besson {\it et al.}  [CLEO Collaboration],
Phys.\ Rev.\ D {\bf 68} (2003) 032002;
%
E. W. Vaandering   [FOCUS  Collaboration], arXiv:hep-ex/0406044;
B.~Aubert {\it et al.}  [BABAR Collaboration],
Phys. Rev.  {\bf D69} (2004) 031101;
%
P.~Krokovny {\it et al.} [Belle  Collaboration], Phys. Rev. Lett.
{\bf 91} (2003) 262002; P.~Krokovny  [Belle Collaboration],
arXiv:hep-ex/0310053.


\bibitem{Colangelo:2004vu}
  For a review see: P.~Colangelo, F.~De Fazio and R.~Ferrandes,
  Mod.\ Phys.\ Lett.\ A {\bf 19} (2004) 2083.


\bibitem{PDG}
S.~Eidelman {\it et al.}  [Particle Data Group],
  Phys.\ Lett.\ B {\bf 592} (2004) 1.

\bibitem{Bonvicini:2005ci}
  G.~Bonvicini {\it et al.}  [CLEO Collaboration],
  arXiv:hep-ex/0510034.

\bibitem{Bardeen:2003kt}
W.~A.~Bardeen, E.~J.~Eichten and C.~T.~Hill,
Phys. Rev.  {\bf D68} (2003) 054024.
  M.~A.~Nowak, M.~Rho and I.~Zahed,
  Acta Phys.\ Polon.\ B {\bf 35} (2004) 2377.


\bibitem{Jaffe:2005sq}
  R.~L.~Jaffe, D.~Pirjol and A.~Scardicchio,
  arXiv:hep-ph/0511081.

\bibitem{Colangelo:2003vg}
  P.~Colangelo and F.~De Fazio,
  Phys.\ Lett.\ B {\bf 570} (2003) 180.

\bibitem{CDF}
  I.~V.~Gorelov  [CDF Collaboration],
  J.\ Phys.\ Conf.\ Ser.\  {\bf 9} (2005) 67.

\bibitem{Colangelo:1995ph}
  P.~Colangelo, F.~De Fazio, G.~Nardulli, N.~Di Bartolomeo and R.~Gatto,
  Phys.\ Rev.\ D {\bf 52} (1995) 6422;
P.~Colangelo and F.~De Fazio,
  Eur.\ Phys.\ J.\ C {\bf 4} (1998) 503.

\bibitem{Abe:2005xj}
  K.~Abe {\it et al.}  [Belle Collaboration],
  arXiv:hep-ex/0507030.

\bibitem{Avery:1994yc}
  P.~Avery {\it et al.}  [CLEO Collaboration],
  Phys.\ Lett.\ B {\bf 331} (1994) 236
  [Erratum-ibid.\ B {\bf 342} (1995) 453].




\end{thebibliography}
\end{document}